\documentstyle[11pt,newpasp,twoside,epsf]{article}
\def\edcomment#1{\iffalse\marginpar{\raggedright\sl#1\/}\else\relax\fi}
\reversemarginpar

\begin{document}
\title{Cold Molecular Gas as a Possible Component
of Dark Matter in the Outer Parts of Disk Galaxies}
\author{ Ronald J. Allen and Rosa Diaz-Miller}

\affil{Space Telescope Science Institute, Baltimore, MD 21218, USA}

\begin{abstract}
In the last few years new evidence has been presented
for the presence of ongoing massive star formation in
the outer HI disks of galaxies. These discoveries
strongly suggest that precursor molecular gas must also
be present in some physical state which is escaping
detection by the usual means (CO(1-0), IR, etc.).  We
present a model for such a gas in a framework which views the HI as
the result of an ongoing ``photodissociation $\Leftrightarrow$
dust grain reformation'' equilibrium in a cold, clumpy
molecular medium with a small area filling factor.
\end{abstract}

\section{Young stars and HI in the outer parts of NGC 6822}

Recently Komiyama et al.\ (2003) and De Blok \& Walter (2003)
have reported the detection of young stars in the outer HI
disk of the nearest dwarf irregular galaxy NGC 6822 ($D \approx 490$ kpc) at
galactocentric radii well beyond $R_{25}$.  These authors discuss the results
in terms of the commonly-accepted picture, where young massive stars form out
of the observed HI in response to some ``triggering'' mechanism such as the
passage of a companion galaxy (in this case a massive intergalactic HI cloud).
However, an ``interaction'' scenario is problematic for NGC 6822, since as De
Blok \& Walter point out there is little evidence for tidal ``impact trauma'' on the
distribution of old stars in the galaxy.

\section{Young stars in the outer disk of M31}

Cuillandre et.  al.\ (2001) discovered a whole population of
upper-main-sequence (mostly B) stars in a large field ($28' \times 28'$)
in the outer parts of M31 ($23 \leq
R_G \leq 33$ kpc at $D \approx 720$ kpc) beyond $R_{25} (\approx 21$ kpc).
Reddening of the accompanying (and much more numerous) red giant stars implied
the presence of dust.  Both the young stars and the reddening correlated well in
position with the brightness of the HI on a scale of $\sim 4' \approx 1$ kpc.
The dust/HI ratio is $\geq 1/3$ of the Galactic value near the sun. The HI
distribution appears quiescent, showing the general rotation of the
galaxy.

\section{An ``Inverse'' Explanation}

If the current view that star formation occurs in molecular clouds is correct, then
the presence of dust and young stars in the outer parts of galaxies
implies that molecular gas must also be
present in these regions. In that case an ``inverse'' explanation for the close
association of HI and young stars is suggested, namely, that
 \textbf{\textit{the young stars have
themselves produced the HI by photodissociation of their parent molecular
clouds}.} The dissociated HI re-forms to H$_2$ on dust grains, and the
observed HI column
density is a result of a dissociation $\Leftrightarrow$ reformation equilibrium. In this case
the HI column density can be computed from (cf. Allen et al. (2003) and
references given there):
\begin{displaymath}
N(HI)= {{7.8 \times 10^{20}} \over {(\delta/\delta_0)}}
\ln{\left[ {{106G_0} \over {n}}
{{\left( {{\delta} \over {\delta_0}} \right)}^{-1/2}}
+ 1 \right]}  {\rm ~cm}^{-2}
\end{displaymath}
\noindent where $n = n(HI) + 2n(H_2)$. $G_0$ and $\delta/\delta_0$ are the FUV
flux and dust/gas ratio w.r.t.\ the Galaxy near the sun .
For instance, $N(HI) \approx 1 \times 10^{20}$ on the surface of
a low-density ($n \approx 10$ cm$^{-3}$) low-metallicity ($\delta/\delta_0
\approx 0.2$) molecular cloud can be maintained by a FUV flux of $G_0 \approx
0.002$, the equivalent of $\sim$ one B3V star every $\sim 100$ pc in the disk.
The re-formation time on the surfaces of dust grains in such a
medium seems long, $\approx 10^8$ yrs, but in fact this is short
($\approx 10\%$)  compared to the rotation time, which is the
relevant time scale here in the absence of galactic collisions.
An equilibrium is therefore likely even under these
extreme conditions of low molecular gas density and low FUV flux.
The accompanying molecular clouds will be cold and of low density,
rendering the normal molecular tracers (e.g. CO(1-0)) undetectable.
More precise calculations are being carried out for M31 where the data of
Cuillandre et.al.\ provides a complete census of all
upper-main-sequence stars (Diaz-Miller, Allen, \& Cuillandre, in preparation).
Initial results indicate that most, and perhaps all, of the HI in the
outer parts of disk galaxies can be accounted for in this way.

\section{Conclusions}

\begin{itemize}
\item \textit{\textbf{The hypothesis:} An $H_2$
photodissociation $\Leftrightarrow$ reformation equilibrium scenario provides a simple
explanation for the coexistence of young stars, dust, and HI in the
outer parts of disk galaxies.}
\item \textit{\textbf{The hypothesis is falsifiable:}  If we find
dynamically quiescent HI regions in the outer parts of galaxies
without dissociating stars ($M(V) \leq -1$, B5 or brighter,
$M \geq 6 M_{\sun}$) then this idea is wrong.}
\item  \textit{\textbf{A prediction:} Even low surface brightness
galaxies such as NGC 2915 will have dissociating B-stars mixed
in with their HI.}
\end{itemize}

\end{document}